# Clinical Translation of Brain-Computer Interface in China: A Landscape Analysis of Investigator-Initiated Trials, Registered Clinical Trials, and Regulatory Approval

Long Chen[#1], Wanyi Qing[#1], Fa Lin[2,3], Lifen Mo[1,4], Xiaoke Chai[5], Wenting Li[6], Zhijie Zhao[7], Zhenhua Song[8], Fengyan Liang*[1], Ming Yin*[1], Yi Yang*[3,9], Jizong Zhao[9]

[1]State Key Laboratory of Digital Medical Engineering, Key Laboratory of Biomedical Engineering of Hainan Province, School of Biomedical Engineering, Hainan University, Sanya, Hainan, China.
[2]Department of Neurosurgery, Beijing Tiantan Hospital, Capital Medical University, Beijing, 100070, China.
[3]China National Clinical Research Center for Neurological Diseases, Beijing, 100070, China.
[4]School of Information and Communication Engineering, Hainan University, Haikou 570000, Hainan, China.
[5]Brain Computer Interface Transitional Research Center, Beijing Tiantan Hospital, Capital Medical University, Beijing, China.
[6]W. P. Carey School of Business, Arizona State University, Tempe, AZ, USA 85287.
[7]Department of Plastic and Reconstructive Surgery, Shanghai Ninth People's Hospital, Shanghai Jiao Tong University School of Medicine, Shanghai, China.
[8]The First Affiliated Hospital of Hainan University, Hainan University, Haikou, Hainan, China.
[9]Beijing Tiantan Hospital, Capital Medical University, Beijing, China.
Correspondence to: Fengyan Liang (fyliang@hainanu.edu.cn), Ming Yin (ming_yin@hainanu.edu.cn), Yi Yang (yangyi_81nk@163.com)

**Abstract**

Neurological injury affects hundreds of millions of people worldwide, yet the loss of motor or communication functions resulting from stroke, spinal cord injury, and neurodegenerative disease remains largely irreversible with existing therapies. Brain–computer interfaces (BCIs) offer a promising pathway for restoring these functions by decoding neural activity into commands that control an external device. Here, we present the first quantitative analysis of China's BCI translational ecosystem, integrating evidence from three critical pillars: investigator-initiated trials (IITs), registered clinical trials, and regulatory-approved products. We systematically analyzed 134 clinical trials from the Chinese Clinical Trial Registry (ChiCTR), 26 IITs, and five BCI-related products approved by the National Medical Products Administration as of June 2026. Results demonstrate that clinical trial registration has increased rapidly since 2020, with geographically clustered yet nationally distributed research centers concentrated primarily in Guangdong, Shanghai, and Jiangsu. Non-invasive systems predominated, accounting for 79.1% of registered studies, with stroke rehabilitation as the leading indication (65.0%). As of June 2026, five BCI-related products received regulatory approvals, including the world's first approved semi-invasive implantable BCI, an invasive closed-loop deep brain stimulation system with real-time local field potential recording, and three non-invasive EEG-based rehabilitation systems. Collectively, these findings characterize a rapidly expanding BCI translational pipeline in China, spanning from early clinical research to regulatory approval. However, long-term implant stability, standardization of clinical infrastructure and workflows, and generalizability of decoding algorithms remain critical barriers to widespread clinical adoption. Addressing these challenges will be essential for integrating BCI technologies into routine clinical practice.

## 1. Introduction

Over the past decade, brain-computer interface (BCI) technologies have shown substantial potential in rehabilitation and medical applications. By decoding neural activity in real time, BCIs enable direct communication between the brain and external devices, bypassing damaged neural pathways and partially restoring lost motor, sensory, or communication functions [1,2]. Recent studies have extended the scope of BCIs beyond restoration, with emerging directions such as brain-to-brain coupling and electroencephalograph-based emotion recognition, reflecting the growing convergence of neural engineering, robotics, and clinical neuroscience [3,4]. Depending on signal acquisition modality, BCIs can be broadly classified into five categories: non-invasive systems using electroencephalography (EEG) and functional near-infrared spectroscopy (fNIRS); semi-invasive electrocorticography (ECoG) placed epidurally or subdurally; intracortical microelectrode arrays; endovascular BCIs; and closed-loop neuromodulation using deep brain stimulation (DBS), responsive neurostimulation (RNS), and stimulation-assisted rehabilitation. Internationally, early clinical studies, including those from the BrainGate consortium, demonstrated that patients with tetraplegia could use robotic limbs and communication systems through decoded neural signals [5]. Long-term follow-up studies from BrainGate demonstrated that intracortical microelectrode arrays can support clinically meaningful neural recordings across multiple trial participants, while also underscoring the progressive decline in signal quality over time as a major barrier to clinical translation [6]. Meanwhile, endovascular BCI studies, such as the Synchron Stentrode program, have further demonstrated the feasibility of stable recordings without craniotomy in participants with severe paralysis [7]. As indicated in a recent review of implantable BCI clinical trials, there is a global shift from single-participant feasibility studies toward structured early feasibility and pivotal trial pathways, with safety, durability, and trial transparency becoming decisive benchmarks for clinical adoption [8].

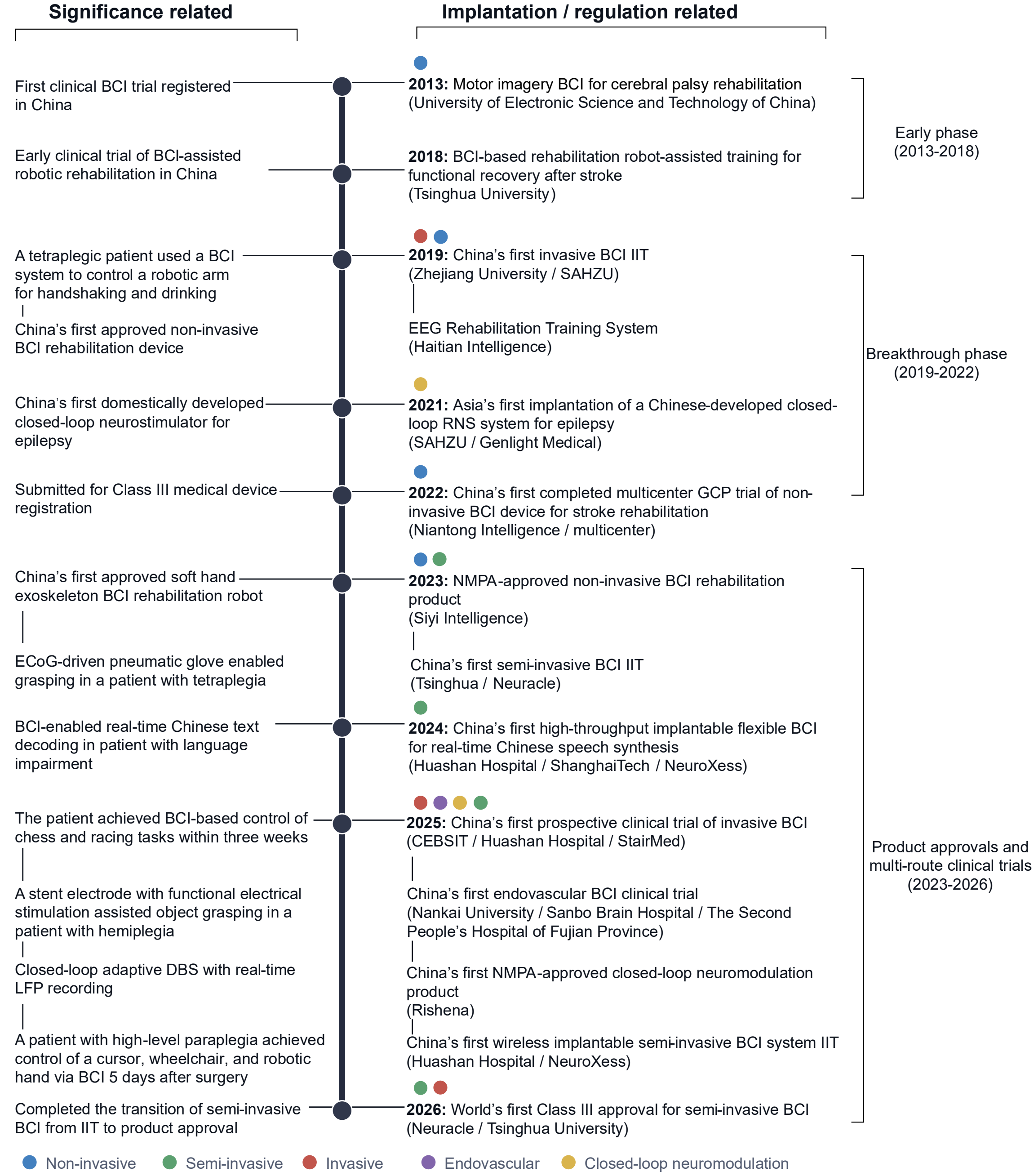

**Fig. 1.** Key milestones in the development of Chinese BCI clinical research, illustrated as a chronological timeline from 2012 to 2026. Events are color-coded by BCI types: blue, non-invasive; green, semi-invasive; red, invasive; purple, endovascular; yellow, closed-loop neuromodulation.

China's clinical BCI trajectory has evolved rapidly over the past decade and can be divided into three distinct phases, as shown in Fig. 1. The foundational phase (2013–2018) was characterized by non-invasive system validation: the first BCI clinical trial was formally registered in China in 2013 by the University of Electronic Science and Technology of China, and by 2015, early closed-loop neuromodulation investigator-initiated trials (IITs) had begun. The breakthrough phase (2019–2022) began with the landmark study in 2019 at Zhejiang University and Second Affiliated Hospital of Zhejiang University School of Medicine (SAHZU), in which a tetraplegic patient implanted with a Utah microelectrode array controlled a robotic arm to drink, making the first invasive BCI IIT in China. This was followed by the National Medical Products Administration (NMPA) approval of Haitian Intelligence's EEG Rehabilitation Training System as China's first certified non-invasive BCI medical device. The current phase (2023–present) has been characterized by accelerated regulatory maturation and technological diversification. Key milestones include China's first domestically developed semi-invasive BCI implantation in 2023 (NEO; Neuracle), the NMPA approval of a closed-loop neurostimulation system in 2025 (Neurestora; Rishena), and the initiation of human trials of China's first fully wireless implanted BCI system, which use a 256-channel flexible ECoG interface (NeuroXess BCI system; NeuroXess Technology). Together, these developments mark China's transition from isolated feasibility demonstrations toward a structured, multi-route clinical translation ecosystem.

The significant expansion of China's BCI field has been driven largely by national policy initiatives. The 14th Five-Year National Key Research and Development Program designated BCI as a priority within the Brain Science and Brain-Inspired Research Initiative, thereby providing substantial translational funding. The Center for Medical Device Evaluation (CMDE) released the Technical Review Points for Implantable Neurostimulators (Trial Version), specifying core requirements for biocompatibility, preclinical performance, and long-term implant safety [9]. In parallel, NMPA introduced the Innovative Medical Device Special Review Procedure, enabling eligible BCI-related devices to enter an accelerated regulatory pathway. These policy instruments have underpinned the gradual formation of a three-tier clinical translation hierarchy spanning IITs, clinical trial registrations, and NMPA-approved products. Existing reviews have primarily focused on individual technologies or specific application domains and are largely based on data collected before the recent acceleration phase. As a result, the three tiers of China's BCI translational pathway have not been examined within a unified analytical framework, and major regulatory and clinical advances achieved between 2023 and 2026 remain insufficiently synthesized. Therefore, in this review, we provide a comprehensive quantitative analysis of China's BCI clinical translation ecosystem as of June 2026, encompassing all five technology approaches, nine indication categories, and the emerging commercialization pipeline.

## 2. Methods

Data were collected from three independent sources. Registered clinical trials were identified through the Chinese Clinical Trial Registry (ChiCTR) using predefined search terms, including "brain-computer interface", "closed-loop neuromodulation", "motor imagery", "functional restoration", and "robotic arm control" (last accessed on June 30, 2026). Studies involving animals or healthy participants were excluded, resulting in 134 eligible registered clinical trials (Supplementary Table 1). Information on IITs was compiled from PubMed, CNKI, institutional websites, and official reports. Gray literature from institutional websites and official reports was used to supplement peer-reviewed publications when publicly available information was incomplete. Eligible IITs were required to involve human clinical investigation of BCI technologies. Animal studies, healthy volunteer studies, conference abstracts lacking sufficient methodological details, and duplicate reports were excluded, yielding 26 IITs (Supplementary Table 2). Information on approved products was obtained from the NMPA Medical Device Database (n = 5). Across all sources, studies involving open-loop neuromodulation without neural signal acquisition, decoding, or feedback components were excluded. Two researchers independently classified technology types according to signal acquisition modality and resolved disagreements by consensus. Clinical indications were assigned using the International Classification of Diseases, Eleventh Revision (ICD-11) framework, with each study counted once under its primary indication. Descriptive statistics were used to summarize temporal trends, geographical distribution, technology tracks, clinical indications, and regulatory status.

On June 30, 2026, the NMPA issued Announcement No. 24 of 2026 [10], which released the Guidance Principle for the Classification and Definition of Brain-Computer Interface Medical Devices and the Guidance Principle for the Generic Naming of Brain-Computer Interface Medical Devices. According to the classification guidance, as shown in Table 1, a product is considered a BCI medical device only when it simultaneously meets four core technical characteristics. First, it must acquire neural signals from the central nervous system (CNS), including the brain, spinal cord, or both, through invasive or non-invasive approaches; products that acquire only peripheral nerve signals, electromyographic (EMG) activity, electrocardiographic (ECG) signals, or other non-CNS bioelectrical signals are not included. Second, it must decode the acquired neural signals in real time, and the decoded commands must serve as the primary operating mechanism of the device, such as controlling an external assistive device, modulating stimulation energy, or triggering neurostimulation. Third, it must enable real-time interaction, including closed-loop feedback where applicable, between the patient and an external assistive or therapeutic device. Fourth, it must be intended to achieve clinically relevant outcomes related to the improvement, restoration, compensation, or replacement of impaired CNS functions. The guidance further places BCI medical devices within China's risk-based medical device classification system. Products with moderate risk are generally regulated as Class II medical devices, for which standard regulatory controls are required to ensure safety and effectiveness. Products with high risk, particularly those involving invasive interfaces, long-term implantation, direct interaction with CNS tissue, active therapeutic intervention, or substantial consequences in the event of malfunction, are generally regulated as Class III medical devices and are subject to more stringent regulatory controls. In the present analysis, these regulatory criteria were used as the operational definition of BCI medical devices and informed the inclusion and exclusion decisions made during data extraction.

**Table 1.** Classification and definition criteria for BCI medical devices

| Criteria | Qualifying BCI Medical Devices | Non-qualifying BCI Medical Devices |
|---|---|---|
| **Signal source** | Acquire CNS neural signals from the brain and/or spinal cord | Acquire only EMG, ECG, peripheral nerve, or other non-CNS signals |
| **Signal decoding** | Decode CNS neural signals in real time and use the decoded neural commands as the primary operating mechanism | Do not decode CNS neural signals in real time or do not use decoded neural commands as the primary operating mechanism |
| **Interaction mode** | Enable real-time patient-device interaction, including closed-loop feedback when applicable | Lack real-time interaction or provide only offline or passive analysis |
| **Clinical purpose** | Improve, restore, compensate for, or replace impaired CNS functions | Intended for entertainment, research display, or non-CNS-related purposes |

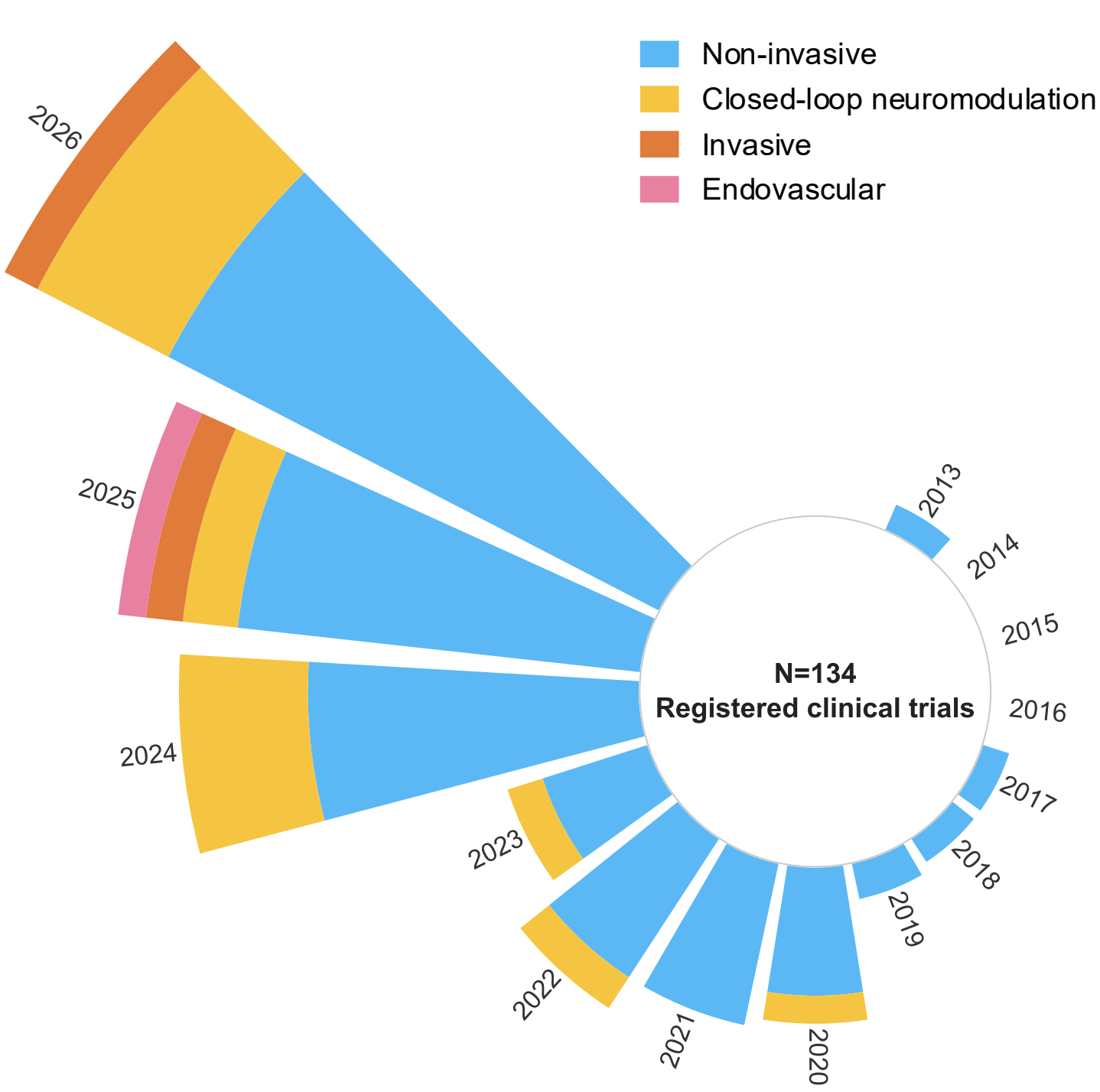

**Fig. 2.** Annual trends of registered BCI clinical trials in China (2013-2026), displayed as a polar area (radial bar) chart. Each angular sector corresponds to one calendar year; the sector area is proportional to the number of newly registered studies (total n = 134). Colors denote technology route: blue, non-invasive; orange, invasive; yellow, closed-loop neuromodulation; pink, endovascular.

## 3. Overall Landscape

China's BCI clinical research follows a three-tier translational framework consisting of: IITs, registered clinical trials, and NMPA-approved products. IITs typically enroll 1–5 participants and focus on safety evaluation and proof-of-concept validation; registered clinical trials provide protocolized clinical evidence and, for device development programs, may support subsequent NMPA submissions; NMPA-approved products represent the final stage of clinical translation. Clinical trial registrations were sparse from 2013 to 2019 (n = 5); rapid growth began in 2020 and accelerated substantially, with 25 new registrations in 2024, 28 in 2025 and 40 registrations recorded by the June 2026 study cutoff, bringing the cumulative total to 134 (Fig.2). This acceleration coincided with the launch of China's Brain Science and Brain-Inspired Intelligence initiative, alongside rising demand for neurorehabilitation services, and the establishment of expedited regulatory review pathways by the NMPA. Geographically, Guangdong (27 studies, 20.1%), Shanghai (21 studies, 15.7%), and Jiangsu (21 studies, 15.7%) together account for more than half of national activity, forming a multipolar research landscape dominated by South and East China (Fig. 3).

## 4. Technological Approaches

### *4.1 Non-Invasive BCI*

Non-invasive BCIs acquire brain signals via EEG or fNIRS, without surgical intervention, and represent the dominant technology track in Chinese clinical trial registrations, accounting for 106 of 134 studies (79.1%; Table 2). The prevailing paradigm couples motor imagery (MI)-BCI with functional electrical stimulation (FES) or robotic exoskeletons within a rehabilitation framework, targeting post-stroke upper and lower-limb recovery through closed-loop sensorimotor feedback and the promotion of cortical plasticity. MI-BCI has been widely investigated as a neurorehabilitation strategy for post-stroke motor recovery [11]. A meta-analysis of eight randomized controlled trials involving 235 participants demonstrated that MI-BCI interventions significantly improved Fugl-Meyer Assessment upper-extremity scores compared with control groups [12]. Nevertheless, the overall certainty of the evidence remains low to moderate owing to limited sample sizes and considerable heterogeneity across outcome measures. Among representative translational programs, the eCon-Hand system (Niantong Intelligence Technology) completed a randomized clinical study involving 168 patients with upper-limb dysfunction after stroke and has advanced to NMPA Class III registration review. In 2023, Siyi Intelligent Technology's BCI rehabilitation training system received NMPA approval as China's first soft hand exoskeleton BCI rehabilitation robot and was registered as a Class II medical device (Table 3). Despite these clinical and regulatory advances, the performance and scalability of non-invasive BCI systems remain constrained by several technical factors. The principal technical limitations of non-invasive BCI systems include the inherently low spatial resolution of scalp EEG and signal degradation due to electrode-skin impedance variations during prolonged use.

### *4.2 Semi-Invasive BCI*

Semi-invasive BCIs position electrodes epidurally or subdurally to acquire ECoG signals, offering higher signal fidelity than non-invasive systems while requiring only moderate surgical burden. No registered semi-invasive BCI clinical trials were identified in the ChiCTR; consequently, semi-invasive systems were not represented as a separate category in Fig. 2. Nevertheless, semi-invasive BCI is among the most advanced IIT-to-product pathways in China (Tables 3, 4). The NEO system, developed jointly by Neuracle and Tsinghua University, employs wireless epidural ECoG electrodes. In October 2023, NEO was successfully implanted at Xuanwu Hospital of Capital Medical University in Beijing, marking the first human implantation of a domestically developed, wireless, fully implantable semi-invasive (epidural) BCI system. Following three weeks of training, the participant with SCI-related tetraplegia achieved decoding accuracies exceeding 90% and successfully performed activities of daily living, including drinking and handshaking [13]. Following completion of a 4-participant feasibility study and a subsequent 32-participant multicenter Good Clinical Practice (GCP) compliant registration trials conducted across 11 hospitals, NEO received NMPA Class III medical device approval on 13 March 2026. To our knowledge, this represents the first regulatory approval of a semi-invasive BCI worldwide. The device is indicated for hand motor compensation in patients with high cervical spinal cord injury (C2-C6) (Table 3). Concurrently, NeuroXess is developing both subdural ECoG and intracortical BCI platforms. Its 256-channel subdural ECoG platform, developed in collaboration with Huashan Hospital, enabled real-time Mandarin speech decoding in 2024 [14]. In December 2025, NeuroXess completed its first battery-powered semi-invasive BCI in a participant with high-level paraplegia, featuring a lead less design without percutaneous connectors (Fig 1). By the fifth postoperative day, the participant successfully achieved cursor control and wheelchair navigation via neural decoding. Meanwhile, NeuraCom's semi-invasive BCI product (FIB2000) has completed long-term implantation in three participants, enabling ECoG-based control of a rehabilitation glove for grasping in a patient with post-stroke hemiplegia and supporting lesion localization in patients with brain tumors. Despite these advances, long-term implant durability, infection risk, and chronic signal degradation remain important challenges for broader clinical adoption.

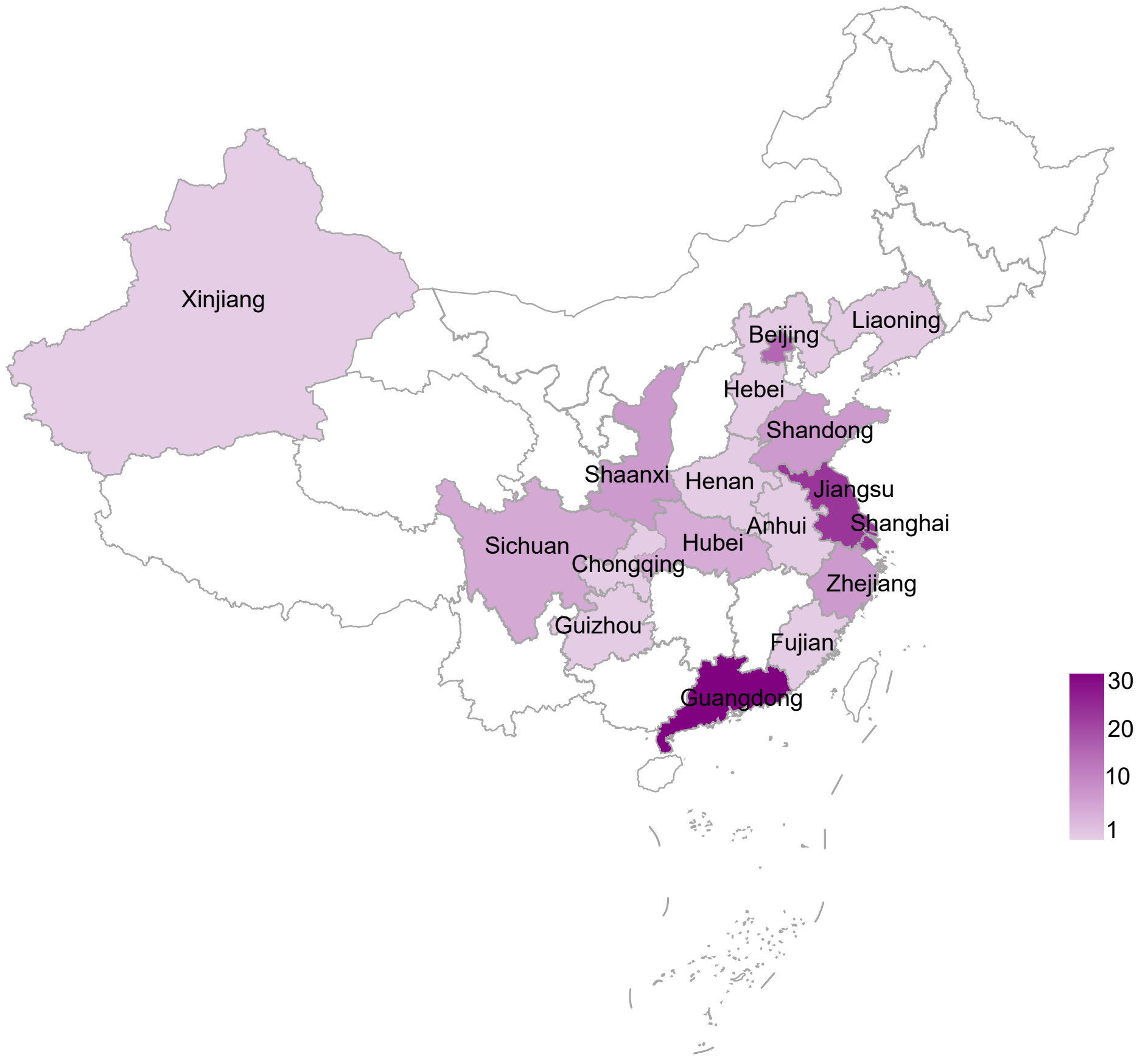

**Fig. 3.** Geographic distribution of registered BCI clinical trials across China. Color intensity (purple gradient, scale: 1-30) represents the number of registered studies per province.

**Table 2.** Registered BCI clinical trials in China by invasiveness

| Invasiveness | n | percentage | Signal Acquisition | Main Indications | Annual Trend |
|---|---|---|---|---|---|
| Non-invasive | 106 | 79.1% | EEG/fNIRS; no-surgical; wearable | Stroke (70); SCI (9); psychiatric & behavioral (10); others (17) | Rapid growth from 2020; 28 studies (2025) → 40 (2026) |
| Invasive | 4 | 3.0% | Intracortical electrode arrays; requires craniotomy | SCI (2); epilepsy (2) | First invasive registered clinical trial in 2025 |
| Endovascular | 1 | 0.7% | Intravascular stent electrode; no craniotomy required | Stroke (1) | China's first endovascular registered clinical trial |
| Closed-loop neuromodulation | 23 | 17.2% | EEG/MI-BCI or fNIRS; outputs include FES, TMS, tDCS, tVNS, SCS and DBS | Stroke (16); SCI (3); psychiatric & behavioral (2); others (2) | Emerged in 2020; acceleration from 2024 |
| Total | 134 | 100% | — | — | 2013–2026; Marked acceleration from 2020 |

*Note: SCI = spinal cord injury; functional electrical stimulation (FES); repetitive transcranial magnetic stimulation (rTMS); intermittent theta-burst stimulation (iTBS;, transcranial direct current stimulation (tDCS); transcutaneous vagus nerve stimulation (tVNS).*

**Table 3.** NMPA-approved BCI products in China

| No. | Product | Company | Invasiveness | NMPA Class | Approval | Indication | Key Feature |
|---|---|---|---|---|---|---|---|
| 1 | NEO/Implantable BCI Hand Motor Compensation System | Neuracle | Semi-invasive (epidural ECoG) | Class III | 2026 | C2-C6 SCI; tetraplegia hand grasp compensation | World's first approved semi-invasive BCI; epidural micro-implant; wireless power |
| 2 | EEG Rehabilitation Training System | Yunnao Medical | Non-invasive (EEG) | Class II | 2025 | Stroke; motor dysfunction rehabilitation | First BCI medical device approved in Chongqing; Motor imagery EEG decoding |
| 3 | Closed-loop Neurostimulation System | Rishena | Invasive (Closed-loop DBS with LFP sensing) | Class III | 2025 | Parkinson's disease | China's first closed-loop adaptive DBS with real-time LFP recording; rechargeable via wireless charging |
| 4 | BCI Rehabilitation Training and Evaluation System | Siyi Intelligence | Non-invasive (EEG) | Class II | 2023 | Stroke, SCI-related finger joint dysfunction | China's first NMPA-approved soft hand exoskeleton BCI rehabilitation robot |
| 5 | EEG Rehabilitation Training System | Haitian Intelligence | Non-invasive (EEG) | Class II | 2019 | Stroke-related upper and lower limb dysfunction | China's first approved non-invasive BCI rehabilitation medical device |

### *4.3 Invasive BCI*

Invasive BCIs implant microelectrode arrays directly into the cerebral cortex to acquire single-neuron signals with high spatiotemporal resolution, primarily targeting patients with high cervical spinal cord injury, epilepsy, or amyotrophic lateral sclerosis (ALS). Four invasive BCI studies were identified among Chinese clinical trial registrations (3.0%), including two spinal cord injury studies and two epilepsy-related studies (Table 2). In 2019, Zhejiang University reported the first intracortical BCI implantation in China. The BCI system enabled three-dimensional robotic arm control, followed by real-time decoding of handwritten character intentions in 2025. The WRS01 system (CEBSIT and StairMed) employs a flexible intracortical microwire electrode. Following the first implantation in an amputee participant in March 2025, three human trials of the 64-channel WRS01 system were completed later that year. The upgraded 256-channel WRS02 platform was released in late 2025 and subsequently entered human clinical evaluation in early 2026. The relatively limited number of registered invasive studies may reflect both the high procedural risk of craniotomy and NMPA's stringent requirements for long-term implant safety and performance data. Consequently, most domestic teams remain in the IIT stage while accumulating evidence for subsequent GCP-compliant registration trial.

### *4.4 Endovascular BCI*

Endovascular BCIs deploy electrodes within blood vessels adjacent to cortical targets, enabling the recording of cortical signals without a craniotomy. Clinical translation of this technology remains in its early stages in China, with only one registered clinical trial and one IIT (Tables 2 and 4). In 2025, Nankai University, in collaboration with Sanbo Brain Hospital, performed China's first endovascular BCI implantation in a patient with post-stroke hemiplegia. Following implantation, the participant demonstrated recovery of functional grasping and object manipulation. Compared with international endovascular BCI programs such as the Synchron Stentrode system, endovascular BCI development in China remains at an earlier stage of clinical validation. Key challenges include long-term intravascular signal stability, maintenance of vascular patency, electrode endothelialization, and the establishment of standardized multicenter clinical protocols.

**Table 4.** Representative IIT studies in China

| Year | Invasiveness | Leading Institution | BCI Device/System | Indication | n | Key Achievement |
|---|---|---|---|---|---|---|
| 2019 | Invasive | Zhejiang University/SAHZU | Utah array (Blackrock) | Cervical SCI tetraplegia | 1 | China's first invasive BCI; 3D robotic arm (2020) & robotic arm-assisted Chinese character writing (2025) |
| 2025 | Invasive | CEBSIT/Huashan Hospital, Fudan University | WRS01 (StairMed) | Amputation/ Quadriplegia/ Motor dysfunction | 3 | Cursor, chess & wheelchair control; 256-ch upgrade (WRS02) completed 2026 |
| 2023-2025 | Semi-invasive | Multicenter | NEO epidural ECoG (Neuracle) | SCI tetraplegia (C2–C6) | 36 | Wireless signal transmission; Received NMPA Class III registration in March 2026 |
| 2024 | Semi-invasive | Fudan Huashan/ShanghaiTech | 256-ch flexible ECoG (NeuroXess Technology) | Language dysfunction | 1 | First real-time Chinese speech synthesis via BCI [14] |
| 2025 | Semi-invasive | Huashan Hospital, Fudan University | wireless BCI system (NeuroXess Technology) | High-level paraplegia | 1 | China's first fully wireless implanted BCI; Cursor & wheelchair control |
| 2024-2026 | Semi-invasive | CIBR/Xuanwu & Tiantan Hospital | Beinao-1 (NeuCyber) | SCI/Stroke Hemiplegia/ALS | 30 | Over 55,000 hours of safe operation; Launched a GCP-compliant registration trial in March 2026 |
| 2022 | Non-invasive | Multicenter | eCon-Hand EEG (Niantong Intelligence) | Stroke upper-limb rehabilitation | 168 | China's first registered clinical trial for a rehabilitation BCI medical device; entered NMPA regulatory submission |
| 2025 | Endovascular | Nankai University/Fujian Sanbo Brain Hospital | Self-developed mesh stent electrode array | Hemiplegia | 1 | China's first endovascular BCI human trial |
| 2021 | Closed-loop neuromodulation | SAHZU | Epilcure RNS system (GenLight MedTech) | Epilepsy | 1 | Asia's First implantation of a closed-loop neural stimulator; entered NMPA regulatory submission |

*Note: n = number of implanted/enrolled subjects; CEBSIT = Center for Excellence in Brain Science and Intelligence Technology; CIBR = Chinese Institute for Brain Research, Beijing.*

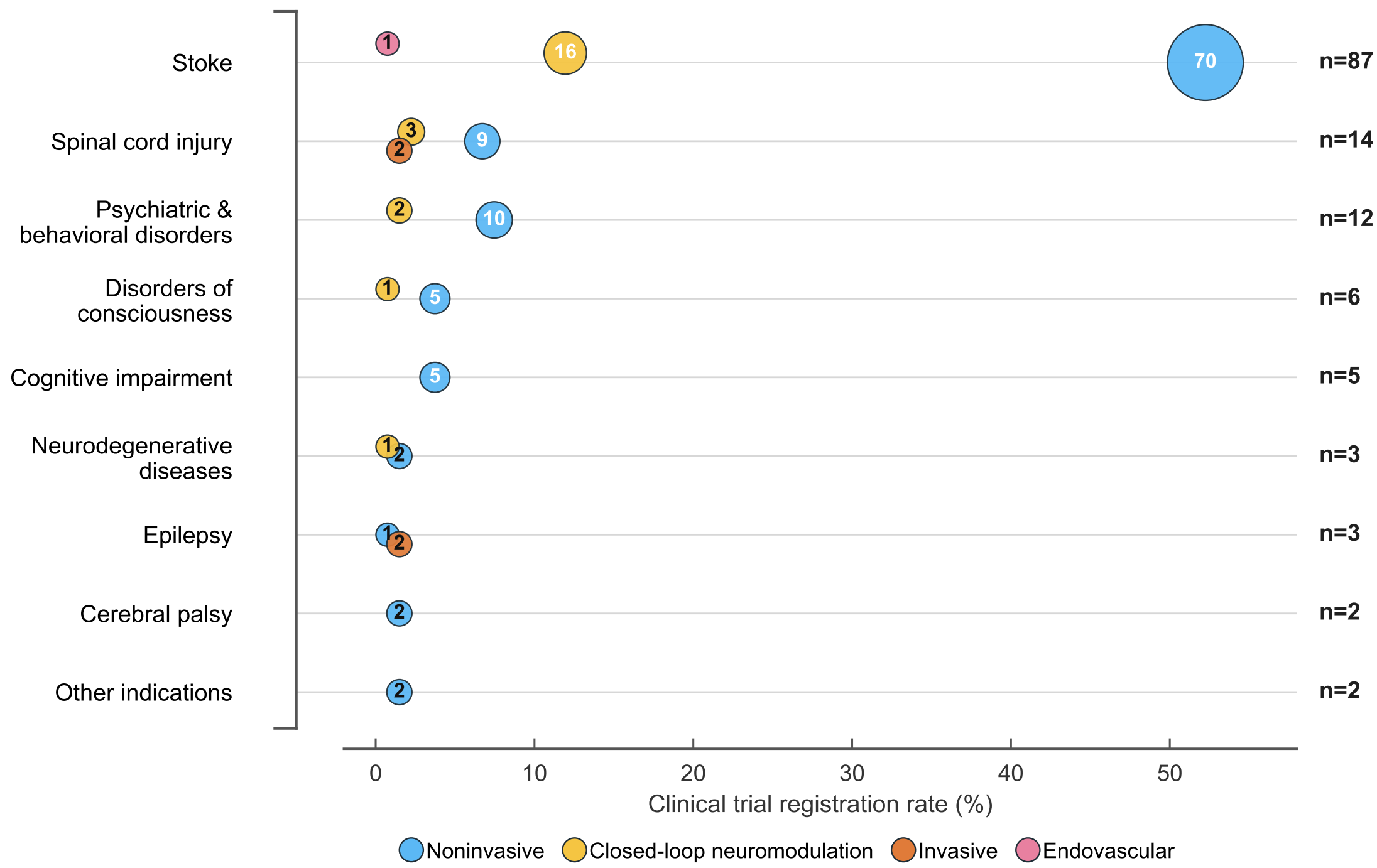

**Fig. 4.** Percentage distribution of ChiCTR-registered (BCI) clinical trials in China by indication and technology route. Each bubble represents one indication–technology combination; bubble area is proportional to trial count (number shown inside); row totals are indicated by k. Colors denote BCI technology type: blue, non-invasive; yellow, closed-loop neuromodulation; orange, invasive; pink, endovascular.

### *4.5 Closed-Loop Neuromodulation*

Closed-loop neuromodulation systems continuously monitor neural activity or behavioral state in real time and deliver adaptive stimulation or peripheral feedback, thereby establishing a bidirectional sensing–stimulation loop that offers greater precision than conventional open-loop paradigms. In the ChiCTR dataset, closed-loop neuromodulation accounted for 23 of 134 registered clinical trials (17.2%), with the first registered study in 2020, followed by rapid expansion after 2024. The main indications include stroke (16 studies), spinal cord injury (3 studies), psychiatric and behavioral disorders (2 studies), and other neuromodulation applications (2 studies) in Table 2. The rehabilitation-oriented subset typically combines EEG or fNIRS-based decoding with feedback modalities such as functional electrical stimulation (FES), repetitive transcranial magnetic stimulation (rTMS), intermittent theta-burst stimulation (iTBS), transcranial direct current stimulation (tDCS), transcutaneous vagus nerve stimulation (tVNS), and spinal cord stimulation. In parallel with advances in rehabilitation, China has made substantial regulatory progress in implantable closed-loop neuromodulation, with Rishena's closed-loop DBS system receiving approval from the NMPA (Table 3). SceneRay's dual-target DBS system is intended for treatment-resistant obsessive-compulsive disorder and opioid addiction relapse prevention. The system combines adaptive sensing with stimulation of the nucleus accumbens (NAc) and the anterior limb of the internal capsule (ALIC), with closed-loop neuromodulation functionality currently under development. In addition, GenLight MedTech's Epilcure responsive neurostimulation system for drug-resistant epilepsy completed Asia's first implantation in 2021 (Table 4), has been implanted in more than 100 patients, and is currently advancing toward NMPA registration.

## 5. Clinical Indication Landscape

The 134 registered clinical trials reveal a dual pattern characterized by strong concentration in a limited number of indications alongside expansion into multiple emerging application areas (Fig. 4 and Table 5). Stroke rehabilitation was by far the predominant indication, accounting for 87 studies (65.0%). The prevailing approach involved MI-EEG-BCI combined with FES or robotic exoskeletons. Among these studies, 70 used non-invasive BCI, 16 employed closed-loop neuromodulation, and one adopted an endovascular approach. Spinal cord injury ranks second (14 studies, 10.4%), spanning non-invasive rehabilitation, closed-loop neuromodulation, and invasive BCI. Psychiatric and behavioral disorders account for 12 studies (9.0%), encompassing depression, addiction, autism spectrum disorder, tinnitus, migraine, motion sickness, and emotional disorders. Disorders of consciousness (6 studies, 4.5%) and cognitive impairment (5 studies, 3.7%) remain at an early stage of clinical investigation. Smaller categories include neurodegenerative diseases (3 studies, 2.2%), epilepsy (3 studies, 2.2%), cerebral palsy (2 studies, 1.5%), and other indications (2 studies, 1.5%). This distribution shows that domestic BCI clinical research is currently anchored in neurorehabilitation, while communication restoration applications, seizure control, psychiatric neuromodulation, and pediatric indications remain comparatively under-represented (Fig. 5).

**Table 5.** Registered BCI clinical trials in China by indication

| Indication | n | Percentage | Invasiveness (n per route) | Scope/Representative Conditions |
|---|---|---|---|---|
| Stroke | 87 | 65.0% | Non-invasive (70) Closed-loop neuromodulation (16) Endovascular (1) | Stroke; Motor rehabilitation; MI-BCI + exoskeleton/FES paradigms |
| Spinal cord injury | 14 | 10.4% | Non-invasive (9) Closed-loop neuromodulation (3) Invasive (2) | Traumatic/Non-traumatic SCI and tetraplegia; Motor and sensory dysfunction, and upper-limb grasping compensation |
| Psychiatric & behavioral | 12 | 9.0% | Non-invasive (10) Closed-loop neuromodulation (2) | Depression, addiction, ASD, tinnitus, migraine, motion sickness and emotion disorders |
| Disorders of consciousness | 6 | 4.5% | Non-invasive (5) Closed-loop neuromodulation (1) | Vegetative state/Minimally conscious state; Neurocritical consciousness assessment and arousal |
| Cognitive impairment | 5 | 3.7% | Non-invasive (5) | Alzheimer's disease and vascular cognitive impairment; Cognitive rehabilitation, screening and monitoring |
| Neurodegenerative diseases | 3 | 2.2% | Non-invasive (2) Closed-loop neuromodulation (1) | Parkinson's disease, ALS and MND; Motor assistance and augmentative communication |
| Epilepsy | 3 | 2.2% | Non-invasive (1) Invasive (2) | Drug-resistant epilepsy and epilepsy-related cognitive assessment |
| Cerebral palsy | 2 | 1.5% | Non-invasive (2) | Pediatric and adult cerebral palsy; Motor rehabilitation via BCI-assisted training |
| Other | 2 | 1.5% | Non-invasive (2) | Rotator cuff injury and lower-limb motor dysfunction; Conditions outside the categories above |
| Total | 134 | 100% | - | - |

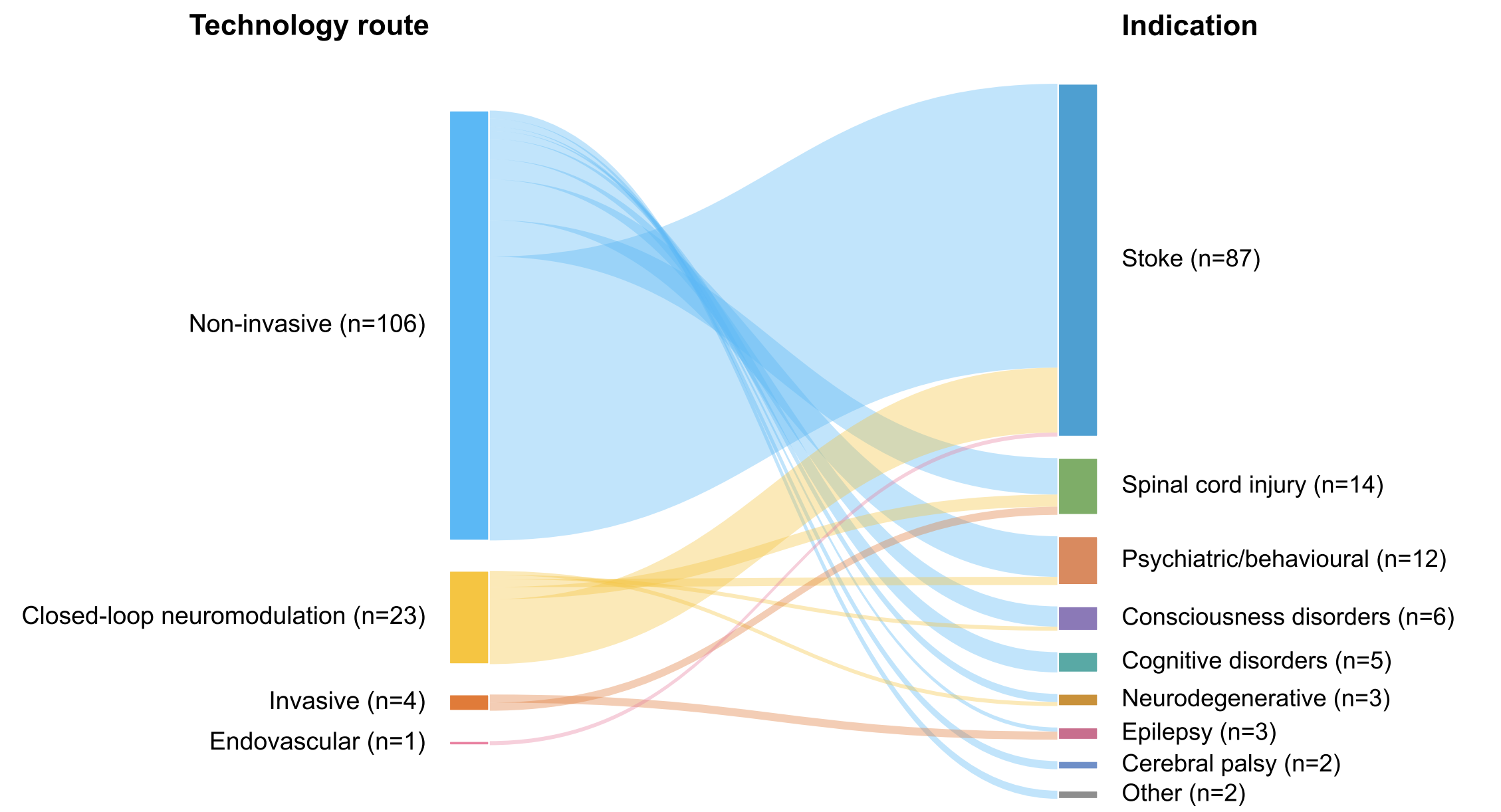

**Fig. 5.** Sankey diagram illustrating the mapping between BCI technology routes (left: non-invasive n = 106, invasive n = 4, closed-loop neuromodulation n = 23, endovascular n = 1) and clinical indications (right) across all 134 registered clinical trials.

**Table 6.** Comparison of BCI clinical translation in China and international programs

| | China | International | Key Comparison |
|---|---|---|---|
| **Clinical scale** | Large registry base; rehabilitation-centered; product-oriented | Smaller implanted cohorts; longer chronic follow-up; feasibility-stage evidence | Scale vs. depth |
| **Technology route** | EEG and fNIRS rehabilitation [15-17]; closed-loop neuromodulation [18,19]; emerging ECoG [14] and endovascular exploration early | Intracortical arrays [5,8,20-23]; ECoG-based speech decoding interfaces [24,25]; endovascular stent-electrodes [26] | Rehabilitation-oriented, lower-risk deployment vs. implant-focused, high-performance neural interfaces |
| **Outcome focus** | Motor recovery; clinical scales; safety and usability; regulatory endpoints | Cursor control; robotic arm; brain-to-text; speech decoding | Clinical recovery vs. neural decoding performance |
| **Translation stage** | IIT-registered trials - NMPA approval | First-in-human trials; long-term feasibility evaluation; limited routine clinical adoption | Approval pathway vs. frontier feasibility |
| **First invasive BCI clinical trial** | 2019; Zhejiang University/SAHZU; Utah array; cervical SCI tetraplegia; robotic arm control | 2004-2006; BrainGate; Utah array; tetraplegia; cursor/external device control | China's first intracortical BCI trial launched much later, while early international work had already validated the feasibility of intracortical microelectrode BCI systems. |
| **Representative implantable BCI companies** | Neuracle, NeuroXess, NeuraCom, Rishena, StairMed, NeuCyber, SceneRay | Neuralink, Synchron, Blackrock Neurotech, Paradromics | China emphasizes clinical rehabilitation and regulatory translation, whereas international programs emphasize high-performance implantable decoding and assistive communication |

## 6. Regulatory Pathway and Commercialization

China adopts a risk-based regulatory framework for BCI medical devices based on invasiveness, intended use, and overall risk. High-risk invasive and implantable BCIs are regulated as Class III medical devices and undergo full technical review by NMPA's CMDE. In contrast, the classification of non-invasive BCIs depends on their intended clinical applications. Simple rehabilitation devices for post-stroke motor recovery are generally classified as Class II devices, whereas systems intended for disease treatment, neuromodulation, or AI-enabled closed-loop compensation are regulated as Class III devices. The 2021 CMDE Technical Review Points for Implantable Neurostimulators (Trial Version) established core requirements for preclinical performance, biocompatibility, and long-term implant safety. Concurrently, the Innovative Medical Device Special Review Procedure has been applied to multiple BCI products. As of June 2026, five products have received NMPA approval (Table 3). For products seeking regulatory approval, registered clinical trials remain a key evidentiary bridge between IIT feasibility work and NMPA review.

Despite substantial progress in regulation and product approval, several challenges continue to hinder the commercialization and widespread clinical adoption of BCI technologies in China. Three major bottlenecks are prominent. First, CMDE requires long-term implant follow-up data, and most invasive studies have not yet met this evidentiary threshold, delaying progression from feasibility work to GCP-compliant registration trial and ultimately to NMPA submission. Second, approved products are not yet broadly reimbursed by the national healthcare insurance system; estimated out-of-pocket costs for implantable BCI procedures may limit widespread adoption and underscore the need for insurance negotiation or payment models based on therapeutic efficacy. Third, implantable BCI procedures require close multidisciplinary collaboration among neurosurgery, neurology, rehabilitation medicine, and biomedical engineering, yet the number of institutions with comprehensive BCI clinical capabilities remains limited.

## 7. Comparison with International Clinical Development

To place China's progress in a global context, it is useful to compare its translational trajectory with those of leading international programs. BCI clinical translation follows different developmental pathways in China and internationally. As shown in Table 6, International studies have accumulated substantial evidence for implantable BCIs, especially chronic intracortical recording, assistive device control, and speech restoration. Between 1998 and 2023, 21 groups conducted 28 implantable BCI trials involving 67 implanted participants worldwide [8]. BrainGate further showed that intracortical decoding can persist for years, although long-term signal quality remains variable [6]. Recent studies from Neuralink, Synchron, and Paradromics indicate renewed momentum in the field, but most implantable BCI trials remain small-scale and investigational. By contrast, China's clinical portfolio is larger in scale and more rehabilitation-oriented. By June 2026, China had 134 registered BCI clinical studies, including 106 non-invasive studies and 87 stroke rehabilitation studies, as well as five NMPA-approved BCI products. This suggests that China has developed comparative strengths in registry scale, rehabilitation-oriented applications, and product translation, In contrast international programs have accumulated more extensive experience in long-term implantable studies, high-performance communication, and control systems. Future Chinese studies should further strengthen long-term safety, long-term signal stability, standardized decoding metrics, and multicenter clinical evidence, especially for invasive and semi-invasive BCI systems.

## 8. Discussion

China has made rapid progress in BCIs, yet several challenges continue to impede their large-scale clinical translation. Major barriers include insufficient long-term implant stability, chronic signal degradation, limited algorithm generalizability, and a lack of standardized BCI clinical infrastructure, including dedicated multidisciplinary wards. In addition, the current clinical landscape remains heavily concentrated on stroke rehabilitation, whereas major indications for invasive BCIs remain underrepresented (Fig. 5).

At the same time, China has developed distinctive industrial and clinical strengths. Large multicenter randomized controlled trials of non-invasive BCIs for motor rehabilitation have generated substantial evidence supporting the clinical translation of rehabilitation-oriented BCI devices. More recently, collaborative efforts by Huashan Hospital and NeuroXess have achieved internationally leading advances in real-time Mandarin tonal speech decoding and Chinese character output using semi-invasive flexible cortical BCIs [14,27], highlighting China's potential in the development of communication neuroprostheses.

However, translating these strengths into internationally recognized clinical evidence will require larger patient cohorts, longer follow-up periods, and more standardized reporting of outcomes. International collaboration will therefore be essential for the next stage of BCI translation. Multicenter studies and harmonized reporting standards could facilitate the generation of long-term safety and efficacy data, enable cross-platform comparisons, and accelerate regulatory translation.

## 9. Limitations of the study

Several limitations of this review should be acknowledged. Because many ongoing studies have not yet reported long-term outcomes, the current landscape remains highly dynamic and may evolve as additional evidence becomes available. In addition, unpublished research programs and proprietary datasets were not included and may therefore be underrepresented, potentially limiting the completeness of the present analysis.

## 10. Conclusion

This study provides the first comprehensive review of the BCI clinical translation in China, linking 26 IITs, 134 registered trials, and 5 NMPA-approved products across five technology tracks. Our analysis shows China's BCI field has undergone a profound transformation, evolving from isolated exploratory studies into a structured translational ecosystem supported by coordinated policies, clinical research, and regulatory innovation. Recent regulatory approvals indicate that multiple BCI platforms are progressing from early feasibility studies toward clinical application.Non-invasive rehabilitation BCIs currently represent China's greatest strength, supported by large-scale clinical studies and an emerging commercialization framework.

In contrast, the long-term signal quality and safety of invasive BCIs remain major challenges that must be addressed before their widespread clinical adoption can be realized. In the coming five years, the next phase of BCI translation in China will depend on progress in three key areas: healthcare reimbursement, standardized clinical infrastructure, and high-quality multicenter clinical evidence. Addressing these challenges will be essential for translating early successes into routine clinical practice and for further strengthening China's leading position in global BCI clinical translation and industrial development.

## 11. Acknowledgments

**Funding:** This work was supported by the National Key Research and Development Program of China (Nos. 2024YFF1400600 and 2024YFF1400602), the Hainan Provincial Natural Science Foundation of China (No.

324RC448), the National Natural Science Foundation of China (No. 32360196), and the R&D Project of Hainan Province (No. ZDYF2025SHFZ023). **Competing interests:** The authors declare that there is no conflict of interest regarding the publication of this article.

**12. Supplementary Materials**

The following supporting materials are available with this article: Supplementary Table 1, China BCI Clinical Studies; Supplementary Table 2, Investigator-Initiated Trials of BCI Technologies in China. All raw supplementary tables for the clinical trials and IITs are deposited on Zenodo, available at https://doi.org/10.5281/zenodo.2127219, and are licensed under the CC BY 4.0 International License.